\documentclass[12pt,a4paper]{article}

\usepackage[utf8]{inputenc}
\usepackage[T1]{fontenc}
\usepackage[american]{babel}
\usepackage{lmodern}
\usepackage{geometry}
\usepackage{amsmath,amssymb}
\usepackage{graphicx}
\usepackage{longtable}
\usepackage{array}
\usepackage{microtype}
\usepackage{setspace}
\usepackage[round,authoryear]{natbib}
\usepackage[hidelinks]{hyperref}
\usepackage{caption}
\usepackage{url}

\graphicspath{{figures/}}

\hypersetup{
    pdftitle={Multifractality in Bitcoin Realized Volatility: Implications for Rough Volatility Modelling},
    pdfauthor={Milan Pontiggia}
}

\begin{document}

\title{Pathwise Roughness of Bitcoin Realized Volatility: Stability Across Time, Sampling, and Volatility Measures}

\author{
Milan Pontiggia\\[0.4em]
\normalsize Master's Student, MAGEFI\\
\normalsize Magist\`ere in International Economics and Finance\\
\normalsize University of Bordeaux, France
}

\date{July 21, 2026}

\maketitle

\begin{abstract}
This paper examines whether Bitcoin realized volatility admits a measurable pathwise roughness index and how stable that estimate is across time and measurement designs. Using one-minute BTC/USD close prices from Bitstamp between 2017 and 2024, realized-volatility paths are constructed at 1-, 5-, 10-, and 15-minute frequencies and evaluated with the model-free normalized \(p\)-variation estimator of \citet{cont2024rough}. A unique root is obtained in 341 of 380 rolling 90-day window-frequency configurations, or 89.7 percent, and in 113 of 128 non-overlapping configurations, or 88.3 percent. Conditional rolling medians of the roughness estimate are 0.054, 0.065, 0.086, and 0.080 at the four respective frequencies, and all finite estimates from the temporal, window-length, and jump-robust analyses are below \(1/2\). Root availability and estimate magnitude nevertheless vary across periods and measurement procedures. Truncation affects root availability primarily at one minute, while bipower variation yields a unique root in 20 of 24 eligible fixed-window configurations. In the eight five-minute fixed-window samples, comparisons with iterative amplitude-adjusted Fourier transform surrogates identify excess MF-DFA width in three samples and excess log-moment curvature in one. Bitcoin realized volatility therefore generally admits a low pathwise roughness estimate, but that estimate is not invariant to time or measurement design. The results concern observed realized volatility and do not directly identify the roughness of latent spot volatility.
\end{abstract}

\clearpage

\section{Introduction}
\label{sec:introduction}

\subsection{Background and Motivation}
\label{subsec:background_motivation}

Volatility is a central input to asset pricing, forecasting, portfolio allocation, and risk management. Its short-scale regularity is therefore relevant to both financial theory and empirical modeling. The rough volatility literature studies volatility paths with local regularity below that of Brownian motion, commonly represented by a Hurst parameter \(H<1/2\). Evidence from equity markets has often placed \(H\) near 0.1, motivating fractional stochastic volatility models that reproduce irregular short-horizon volatility dynamics and several features of option markets \citep{gatheral2018volatility}. Subsequent work has examined how roughness can be estimated from high-frequency data and under which conditions such estimates identify the regularity of an underlying volatility process \citep{fukasawa2022consistent,cont2024rough}.

Empirical roughness estimates also depend on the object being measured. Since spot volatility is latent, studies generally construct realized measures from discrete returns. The resulting paths reflect sampling frequency, market microstructure, measurement error, and discontinuous price movements. A roughness estimate obtained from realized volatility should therefore be interpreted as a property of the observed measure unless additional assumptions connect it to latent spot volatility \citep{cont2024rough}.

Bitcoin provides a useful setting in which to examine these measurement issues. The market operates continuously, trading is distributed across venues, and liquidity conditions have changed substantially over time. Cryptocurrency markets exhibit cross-venue price dispersion and limits to arbitrage \citep{makarov2020trading}, while Bitcoin returns display heavy tails, jumps, volatility clustering, and regime variation \citep{chaim2018jumps,caporale2019regime,lahmiri2018chaos}. These characteristics may affect both the construction of realized volatility and the stability of estimated path regularity. At the same time, minute-level observations permit systematic comparisons across periods, sampling frequencies, and volatility measures.

\subsection{Research Gap and Question}
\label{subsec:research_gap_question}

Most empirical evidence on rough volatility concerns equity and index markets. Evidence for Bitcoin is more limited, particularly regarding the stability of pathwise estimates across time and measurement designs. An estimate from a single interval may be sensitive to the selected period, while aggregation and the treatment of extreme returns may alter both the observed path and the availability of a finite estimate. Existing studies of Bitcoin also report variation in scaling behavior across periods and moment orders \citep{takaishi2020rough,takaishi2025multifractality}. These findings make it useful to distinguish the existence of a pathwise roughness estimate from the magnitude and stability of that estimate.

The paper addresses the following question:

\begin{quote}
\emph{Does Bitcoin realized volatility admit a measurable pathwise roughness index, and how stable is that estimate across time periods, sampling frequencies, window lengths, and volatility constructions?}
\end{quote}

The analysis treats root availability and finite roughness estimates as separate outcomes. A unique normalized \(p\)-variation root provides one estimate for a given path and partition. An absent or nonunique root does not. Conditional on a unique root, the magnitude of the estimated roughness index indicates the regularity assigned to that observed path. This distinction permits variation in estimator availability to be examined independently from variation in finite estimates.

\subsection{Empirical Design and Main Findings}
\label{subsec:empirical_design_findings}

The empirical analysis uses one-minute BTC/USD close prices from Bitstamp between 2017 and 2024. Realized-volatility paths are constructed at 1-, 5-, 10-, and 15-minute frequencies from a common one-minute return base. Pathwise roughness is estimated with the normalized \(p\)-variation statistic of \citet{cont2024rough}. The primary temporal design consists of 90-day windows beginning every 30 days. Every third window forms a non-overlapping subset, while one fixed 90-day window per year forms a diagnostic panel for the jump, structural-break, scaling, and surrogate analyses. Additional comparisons use 60- and 120-day windows, locally truncated realized volatility, and bipower variation.

A unique root is obtained in 341 of 380 rolling window-frequency configurations, or 89.7 percent. The corresponding rate is 113 of 128, or 88.3 percent, in the non-overlapping subset. Because the rolling windows overlap, these proportions describe the observed configurations rather than independent replications. Among rolling configurations with a unique root, median estimates of \(H\) are 0.054, 0.065, 0.086, and 0.080 at the 1-, 5-, 10-, and 15-minute frequencies, respectively. The fixed diagnostic panel yields a unique root in 26 of 32 configurations. All finite estimates produced by the temporal, window-length, and jump-robust analyses are below \(1/2\).

The estimates are sensitive to empirical design. Root availability and the distribution of finite \(H\) vary across periods, frequencies, and window lengths. Truncation reduces root availability most clearly at one minute and changes the magnitude of finite estimates. Bipower variation yields a unique root in 20 of 24 eligible fixed-window configurations, with roots remaining common at aggregated intraday frequencies. Rolling root availability shows no significant monotonic trend from 2017 to 2024, while annual median \(H\) increases at the 10- and 15-minute frequencies. Penalized segmentation also identifies frequent changes in realized-volatility levels.

Scaling diagnostics provide secondary evidence on heterogeneity across moments and temporal scales. In the five-minute fixed-window panel, comparisons with iterative amplitude-adjusted Fourier transform surrogates show that the empirical marginal distribution and approximate linear dependence reproduce much of the observed scaling heterogeneity, with residual differences confined to selected periods. The combined results indicate that Bitcoin realized volatility generally admits a low pathwise roughness estimate, while the availability and magnitude of that estimate vary with time and measurement design.

\subsection{Contributions}
\label{subsec:contributions_structure}

The paper makes three contributions. First, it provides a pathwise assessment of Bitcoin realized-volatility roughness using a model-free normalized \(p\)-variation estimator. This approach evaluates the observed path without specifying a complete stochastic volatility model and reports root existence separately from the value of a finite estimate.

Second, the paper examines temporal and measurement stability using rolling and non-overlapping windows, alternative sample lengths, four sampling frequencies, and jump-robust volatility constructions. This design documents the extent to which the empirical estimate varies with the period and measurement procedure.

Third, the paper evaluates scaling heterogeneity through MF-DFA, log-moment scaling, and wavelet leaders. Detailed comparisons with shuffled, phase-randomized, and iterative amplitude-adjusted Fourier transform surrogates are conducted for the five-minute fixed-window panel. These diagnostics assess whether the observed scaling statistics exceed those reproduced by controls that preserve selected distributional and dependence properties.

\section{Literature Review}
\label{sec:literature_review}

\subsection{Rough Volatility and the Measurement of Roughness}
\label{subsec:rough_volatility_literature}

The rough volatility literature studies volatility processes with sample paths that are less locally regular than Brownian motion. In fractional specifications, this property is represented by a Hurst exponent \(H<1/2\). Using high-frequency realized-volatility series from equity markets, \citet{gatheral2018volatility} report scaling behavior consistent with values of \(H\) near 0.1. This evidence has motivated fractional stochastic volatility models designed to reproduce irregular short-horizon volatility dynamics.

Local roughness is distinct from persistence. Roughness describes path regularity over short scales, whereas persistence concerns the decay of dependence over longer horizons. Fractional Brownian motion links these properties through a single Hurst parameter. In contrast, \citet{bennedsen2022decoupling} develop a volatility framework in which short-term roughness and long-term dependence are governed separately. A low roughness estimate should therefore not be interpreted automatically as evidence of either long memory or its absence.

The empirical interpretation of roughness also depends on the object and method of estimation. Scaling regressions based on volatility increments, parametric estimators for fractional stochastic volatility models, and pathwise variation methods rely on different assumptions and do not necessarily target the same quantity. \citet{fukasawa2022consistent} develop a quasi-likelihood estimator that accounts explicitly for error in realized volatility and establish consistency within a specified fractional stochastic volatility setting. Pathwise methods instead assess the regularity of an observed signal without estimating a complete parametric model.

This distinction is important because spot volatility is latent and empirical studies generally use realized measures constructed from discrete returns. Sampling error and the construction of the realized measure can affect its apparent regularity. \citet{cont2024rough} show that realized volatility may exhibit a roughness estimate below \(1/2\) even when latent spot volatility has diffusive regularity. An estimate based on realized volatility must consequently be interpreted as a property of the observed measure unless additional assumptions connect it to the latent volatility process.

\subsection{The Normalized \texorpdfstring{\(p\)}{p}-Variation Estimator}
\label{subsec:normalized_p_variation_literature}

The normalized \(p\)-variation estimator introduced by \citet{cont2024rough} provides a nonparametric, pathwise measure of regularity. The underlying variation index identifies a critical order \(p\), with the associated roughness index expressed as \(H=1/p\). Unlike estimators derived from a specified stochastic volatility model, this approach does not require a distributional law for the observed path.

The discrete construction of \citet{cont2024rough} divides an observed path into coarse intervals and compares, within each interval, the absolute increment across the coarse endpoints with the sum of absolute powered increments on the corresponding fine partition. The resulting ratios are averaged across intervals using the coarse time increment as the normalizing weight. For a path with a well-defined variation index, the order at which the normalized statistic equals one identifies the empirical variation order and hence the roughness index. A unique crossing supports a single pathwise estimate, whereas an absent or nonunique crossing does not.

Although the estimator is model-free, its finite-sample behavior remains conditional on the observed path and partition. Sample length, measurement error, volatility construction, and the allocation of fine increments to coarse blocks can affect whether a unique crossing is found and can change the magnitude of the resulting estimate. Empirical applications should therefore report root availability separately from the distribution of finite estimates and identify the partition used in estimation. These considerations also reinforce the distinction between the roughness of a realized-volatility series and that of latent spot volatility. The mathematical definitions and empirical implementation are presented in Section~\ref{sec:data_methodology}.

\subsection{Bitcoin Volatility, Jumps, and Temporal Variation}
\label{subsec:bitcoin_volatility_literature}

Several features of Bitcoin trading are relevant to high-frequency volatility measurement. The market operates continuously and trading is distributed across venues with different liquidity conditions. \citet{makarov2020trading} document cross-market price dispersion and limits to arbitrage in cryptocurrency markets. These characteristics provide extensive intraday observations but can also make realized measures sensitive to the selected venue, sampling frequency, and market conditions.

The empirical literature documents heavy tails, jumps, and time variation in Bitcoin returns and volatility. \citet{chaim2018jumps} identify discontinuous movements in Bitcoin returns and volatility, while \citet{caporale2019regime} find that models allowing for asymmetry and regime switching improve the representation of cryptocurrency volatility. \citet{lahmiri2018chaos} report long-range correlations, heavy-tailed distributions, and heterogeneous scaling behavior in Bitcoin prices and returns. This evidence indicates that a roughness estimate obtained from one period or one volatility construction may not describe the full sample.

Jump-robust measures offer a way to examine the sensitivity of roughness estimates to extreme returns. Realized bipower variation was developed to estimate the continuous component of quadratic variation under conditions that allow for jumps \citep{barndorff2004power}. Return truncation provides a related empirical control by reducing the influence of observations that exceed a locally determined threshold. Comparisons of standard realized volatility with truncated measures and bipower variation can therefore show whether root availability and finite roughness estimates are sensitive to the treatment of large returns. Such comparisons do not, by themselves, identify jumps as the cause of any difference.

Prior Bitcoin studies also suggest that measured regularity varies with sampling and moment order. Applying MF-DFA to Bitcoin log-volatility increments, \citet{takaishi2020rough} obtain generalized Hurst exponents below \(1/2\) and report variation across moment orders. Their shuffled-series comparisons indicate that the marginal distribution contributes to the measured scaling heterogeneity. Using Bitcoin realized volatility, \citet{takaishi2025multifractality} report sensitivity of Hurst estimates to the sampling period and variation in scaling heterogeneity over time. These findings motivate an analysis that separates the existence of a pathwise roughness estimate from its stability across periods, frequencies, and volatility measures.

\subsection{Scaling Heterogeneity and Surrogate Analysis}
\label{subsec:scaling_heterogeneity_literature}

Scaling heterogeneity refers to variation in empirical scaling behavior across moment orders or temporal scales. Multifractal detrended fluctuation analysis estimates generalized Hurst exponents from detrended fluctuations and is designed to accommodate polynomial trends \citep{kantelhardt2002multifractal}. Log-moment scaling evaluates whether empirical structure-function exponents depart from linearity in the moment order. Wavelet-leader methods characterize multiscale regularity through local wavelet quantities and associated scaling functions \citep{wendt2007bootstrap}. Because these approaches use different transformations, they provide complementary rather than interchangeable evidence.

Measured scaling heterogeneity may reflect heavy-tailed marginal distributions, temporal dependence, or a combination of both. Surrogate data provide reference series that preserve selected properties of the observed data while altering others \citep{schreiber2000surrogate,desouza2013components}. Comparisons with several surrogate types are therefore more informative than reliance on a single randomized control.

Shuffling preserves the empirical marginal distribution while removing the original temporal ordering. Phase randomization preserves Fourier amplitudes and thus the linear dependence represented by the power spectrum, while changing phase relationships and generally not preserving the empirical marginal distribution. Iterative amplitude-adjusted Fourier transform surrogates approximately preserve both the empirical marginal distribution and the power spectrum. Differences between the observed and surrogate statistics indicate scaling behavior not reproduced under the corresponding controls. They do not provide a unique causal decomposition because surrogate matching is approximate and distributional and dependence features may interact.

Pathwise roughness and scaling heterogeneity describe related but distinct aspects of a volatility series. The normalized \(p\)-variation estimator concerns regularity along a specified sequence of partitions, whereas MF-DFA, log-moment scaling, and wavelet-leader methods examine variation across moments and scales. The latter methods can therefore provide supporting evidence on whether a single scaling description adequately summarizes the observed series. Surrogate comparisons assess whether the diagnostic values exceed those reproduced under controls for the marginal distribution and linear dependence structure.
\section{Data and Methodology}
\label{sec:data_methodology}

\subsection{Mathematical Framework}
\label{subsec:mathematical_framework}

\subsubsection{\texorpdfstring{\(p\)}{p}-Variation, Variation Index, and H\"older Regularity}
\label{subsubsec:p_variation_holder}

Let \(X:[0,T]\rightarrow\mathbb{R}\) be a continuous path and let
\[
    \Pi=\{\pi_n\}_{n\geq 1},
    \qquad
    \pi_n=\{0=t_0^{(n)}<t_1^{(n)}<\cdots<t_{N_n}^{(n)}=T\},
\]
be a refining sequence of partitions with mesh
\[
    |\pi_n|
    =
    \max_{0\leq i<N_n}
    \left(t_{i+1}^{(n)}-t_i^{(n)}\right)
    \longrightarrow 0.
\]
For \(p\geq 1\), the \(p\)-variation sum along \(\pi_n\) up to time \(t\) is
\begin{equation}
    V_p(X;\pi_n,t)
    =
    \sum_{\substack{0\leq i<N_n\\t_{i+1}^{(n)}\leq t}}
    \left|
        X\!\left(t_{i+1}^{(n)}\right)
        -
        X\!\left(t_i^{(n)}\right)
    \right|^p.
    \label{eq:p_variation_sum}
\end{equation}
The path has finite \(p\)-variation along \(\Pi\) when these sums converge for every \(t\) to a finite, continuous, increasing function. Following \citet{cont2024rough}, the variation index is
\begin{equation}
    p_{\Pi}^{\ast}(X)
    =
    \inf
    \left\{
        p\geq 1:
        X \text{ has finite \(p\)-variation along } \Pi
    \right\},
    \label{eq:variation_index}
\end{equation}
and the associated variation-based roughness index is
\begin{equation}
    H_{\Pi}(X)=\frac{1}{p_{\Pi}^{\ast}(X)}.
    \label{eq:pathwise_roughness_index}
\end{equation}

H\"older regularity gives a related local description. A path is pointwise \(\alpha\)-H\"older at \(t\) if constants \(C>0\) and \(\eta>0\) exist such that
\begin{equation}
    |X(u)-X(t)|
    \leq C|u-t|^{\alpha}
    \label{eq:holder_condition}
\end{equation}
whenever \(|u-t|<\eta\). Under homogeneous local scaling and suitable regularity conditions, a H\"older exponent \(H\) corresponds to a variation index \(1/H\). This correspondence need not hold pointwise for an arbitrary path with heterogeneous local regularity. The empirical analysis therefore concerns a variation-based property of an observed path along a specified partition.

\subsubsection{Normalized \texorpdfstring{\(p\)}{p}-Variation}
\label{subsubsec:normalized_p_variation_framework}

The normalized \(p\)-variation statistic of \citet{cont2024rough} estimates the variation index from discrete observations. Let \(X_0,\ldots,X_N\) denote observations on a uniform grid whose time interval has been normalized to \([0,1]\). For \(K\) coarse blocks, define
\begin{equation}
    n=\left\lfloor\frac{N}{K}\right\rfloor,
    \qquad
    L=Kn.
    \label{eq:partition_dimensions}
\end{equation}
The first \(L+1\) observations then form \(K\) equal blocks containing \(n\) fine increments each. The normalized statistic is
\begin{equation}
    W(L,K,p)
    =
    \frac{1}{K}
    \sum_{j=0}^{K-1}
    \frac{
        \left|X_{(j+1)n}-X_{jn}\right|^p
    }{
        \displaystyle
        \sum_{i=jn}^{(j+1)n-1}
        \left|X_{i+1}-X_i\right|^p
    },
    \qquad p>1.
    \label{eq:normalized_p_variation}
\end{equation}
Within each block, the numerator is the powered increment between the coarse endpoints and the denominator is the \(p\)-variation over the constituent fine increments. The factor \(1/K\) is the coarse time increment on the normalized interval.

The empirical variation order \(\widehat{p}\) satisfies
\begin{equation}
    W(L,K,\widehat{p})=1,
    \label{eq:root_condition}
\end{equation}
with corresponding roughness estimate
\begin{equation}
    \widehat{H}=\frac{1}{\widehat{p}}.
    \label{eq:roughness_estimate}
\end{equation}
A unique root yields one finite estimate. If no root exists on the admissible domain, no estimate is assigned. Multiple roots likewise do not identify a unique variation order. Root availability and the magnitude of finite \(\widehat{H}\) are consequently reported as separate outcomes. Both are conditional on the observed path and finite-sample partition.

\subsection{Data and Sampling Design}
\label{subsec:data_sampling_design}

\subsubsection{Data Source and Preprocessing}
\label{subsubsec:data_preprocessing}

The data consist of minute-level BTC/USD observations from the Bitstamp exchange, distributed through the Bitcoin Historical Data archive \citep{zielinski2025bitcoin}. The empirical period extends from 2017 through 2024. The one-minute close price is the sole base input to the return, volatility, roughness, and scaling calculations.

Unix timestamps are converted to UTC, ordered chronologically, and checked for uniqueness and exact minute alignment. Close prices must be finite and strictly positive. Each window is reindexed to a regular one-minute UTC grid, and the close immediately preceding the window is used only to calculate its first return. Prices and returns are not interpolated or imputed.

A one-minute return is available only when both adjacent close prices are observed. An aggregated volatility observation is retained only when all of its constituent one-minute returns are available. Estimation is applied to the longest contiguous finite path if missing observations occur. The configurations included in the rolling, fixed-window, and window-length analyses have complete close-price coverage.

\subsubsection{Sampling Windows}
\label{subsubsec:sampling_windows}

The primary temporal design uses 90-day windows beginning every 30 days from 1 January 2017. A window is retained when its complete 90-day interval falls within the empirical period. This procedure yields 95 windows and 380 window-frequency configurations across the four sampling frequencies. Adjacent windows overlap by 60 days. The resulting proportions and distributions provide descriptive temporal coverage and are not interpreted as estimates from independent samples.

Every third rolling window is retained to form a non-overlapping subset. The subset contains 32 windows and 128 window-frequency configurations. It provides a comparison based on distinct 90-day periods, although the four frequency-specific series within a given window share the same one-minute price base.

A separate diagnostic panel contains one fixed 90-day window for each year from 2017 to 2024. Using common calendar intervals holds the observation period constant across sampling frequencies and diagnostic methods. These windows are selected diagnostic samples and are not treated as representative of their full calendar years. The complete date schedule is reported in Table~\ref{tab:fixed_windows_appendix}.

Sample-horizon sensitivity is examined with 60-, 90-, and 120-day windows beginning on the same eight fixed start dates. Holding the starting date constant separates changes associated with the observation horizon from changes associated with selecting another part of the year.

\subsection{Volatility Measures}
\label{subsec:volatility_measures}

\subsubsection{Primary Realized Volatility}
\label{subsubsec:primary_realized_volatility}

Let \(P_t\) denote the one-minute close price. The one-minute log return is
\begin{equation}
    r_t=\log P_t-\log P_{t-1}.
    \label{eq:one_minute_return}
\end{equation}
For an \(m\)-minute interval \(b\), where \(m\in\{1,5,10,15\}\), realized volatility is defined as
\begin{equation}
    RV_b^{(m)}
    =
    \left(
        \sum_{i=1}^{m} r_{b,i}^{2}
    \right)^{1/2},
    \label{eq:realized_volatility}
\end{equation}
where \(r_{b,i}\) is the \(i\)th one-minute return in the interval. Thus, \(RV_b^{(1)}=|r_b|\). Aggregation bins are aligned with the beginning of each analysis window and retained only when complete. The pathwise estimator is applied directly to realized-volatility levels without a logarithmic transformation.

\subsubsection{Alternative Volatility Measures}
\label{subsubsec:alternative_volatility_measures}

Two alternative constructions assess sensitivity to the volatility measure. First, locally truncated realized volatility reduces the contribution of extreme one-minute returns. Let \(\widehat{\sigma}_t\) be the median of the preceding 1,440 absolute one-minute returns, divided by the Gaussian consistency constant \(\Phi^{-1}(0.75)\). For threshold multiplier \(c\in\{4,6,8\}\),
\begin{equation}
    TRV_b^{(m,c)}
    =
    \left(
        \sum_{i=1}^{m}
        r_{b,i}^{2}
        \mathbf{1}
        \left\{
            |r_{b,i}|
            \leq c\widehat{\sigma}_{b,i}
        \right\}
    \right)^{1/2}.
    \label{eq:truncated_rv}
\end{equation}
Returns above the local threshold contribute zero to the truncated sum. Boundary rules for the trailing scale are reported in Appendix A.

Second, realized bipower variation provides a jump-robust proxy for the continuous component of quadratic variation \citep{barndorff2004power}. Within an \(m\)-minute interval,
\begin{equation}
    BPV_b^{(m)}
    =
    \frac{\pi}{2}
    \sum_{i=2}^{m}
    |r_{b,i}|\,|r_{b,i-1}|,
    \qquad m\in\{5,10,15\}.
    \label{eq:bipower_variation}
\end{equation}
The roughness estimator is applied to \(\sqrt{BPV_b^{(m)}}\). Bipower variation is not defined for the one-minute bins used here because each bin must contain at least two returns. Truncation and bipower variation are measurement comparisons. Differences from the baseline do not identify a unique source for the change in estimated roughness.

\subsection{Pathwise Roughness Estimation}
\label{subsec:pathwise_roughness_estimation}

For each volatility series, \(N\) denotes the number of fine increments in its longest contiguous finite path. The baseline number of coarse blocks is
\begin{equation}
    K=\max\left\{2,\left\lfloor\sqrt{N}\right\rfloor\right\}.
    \label{eq:baseline_k}
\end{equation}
The block size is \(n=\lfloor N/K\rfloor\), and the first \(L=Kn\) increments are used so that every block contains the same number of fine increments. Any unused remainder is excluded from the end of the path.

The search is conducted over \(H\in[0.01,0.95]\), with \(p=1/H\), which maintains the restriction \(p>1\). The statistic is evaluated in logarithmic form for numerical stability, and each detected sign change in \(\log W\) is refined with a bracketing root solver. A finite \(\widehat{H}\) is reported only when exactly one root is detected. Each empirical summary therefore includes both the unique-root rate and the distribution of \(\widehat{H}\) conditional on a unique root.

The numerical implementation was checked on fractional Brownian motion paths with known Hurst parameters; the simulation design and results are reported in Appendix A. Appendix A also reports a limited partition check for the fixed 2021 window.

\subsection{Temporal Stability and Structural Diagnostics}
\label{subsec:temporal_structural_diagnostics}

\subsubsection{Temporal Comparisons and Trends}
\label{subsubsec:temporal_comparisons_trends}

Temporal stability is evaluated by comparing root availability and finite-\(\widehat{H}\) distributions across the rolling and non-overlapping designs. Each fixed-window estimate is also located relative to the interquartile range of rolling estimates whose windows begin in the same calendar year. The 60-, 90-, and 120-day samples provide a separate comparison of root rates, medians, and interquartile ranges across observation horizons.

For temporal trends, the rolling estimates are grouped by start year and frequency to obtain annual unique-root rates and annual median finite \(\widehat{H}\). Monotonic patterns are summarized with the Theil-Sen slope and Spearman rank correlation \citep{sen1968estimates}. Trends in the scaling diagnostics use the eight yearly observations from the fixed diagnostic panel. Because each frequency-specific trend is based on eight annual values, the results are interpreted as descriptive temporal evidence.

\subsubsection{Volatility-Level Variation and Structural Breaks}
\label{subsubsec:volatility_level_breaks}

Changes in realized-volatility levels are characterized for the 32 fixed year-frequency configurations. Penalized binary segmentation with a squared-error cost is applied to each standardized volatility path \citep{truong2020selective}. The procedure selects the number and location of breaks through a sample-size-dependent penalty rather than imposing a fixed number. Detected positions are mapped to UTC timestamps. The minimum segment length, candidate spacing, and penalty specification are documented in Appendix A.

Augmented Dickey-Fuller and Kwiatkowski-Phillips-Schmidt-Shin tests provide supporting level diagnostics \citep{dickey1979distribution,kwiatkowski1992testing}. The augmented Dickey-Fuller test has a unit-root null, while the Kwiatkowski-Phillips-Schmidt-Shin test has a level-stationarity null. Joint support for level stationarity is recorded only when the former null is rejected and the latter is not rejected at the 5 percent level. Lag-selection details and complete results are provided in Appendices A and B. These diagnostics characterize the observed volatility levels; stationarity is not imposed as a condition for the normalized \(p\)-variation estimator.

\subsection{Scaling Heterogeneity and Surrogate Design}
\label{subsec:scaling_surrogate_design}

\subsubsection{Scaling Diagnostics}
\label{subsubsec:scaling_diagnostics}

Scaling heterogeneity is examined with MF-DFA, log-moment scaling, and wavelet leaders. The first two methods use moment orders
\begin{equation}
    q\in\{0.5,1.0,1.5,2.0,2.5,3.0\}
    \label{eq:positive_q_grid}
\end{equation}
and a common target scale range from 60 to 2,880 minutes. Negative moment orders are excluded from these calculations because the realized-volatility paths contain exact zeros.

MF-DFA estimates generalized exponents \(h(q)\) from detrended fluctuations,
\begin{equation}
    F_q(s)\propto s^{h(q)},
    \label{eq:mfdfa_scaling}
\end{equation}
using first-order polynomial detrending \citep{kantelhardt2002multifractal}. The primary summary is the width
\begin{equation}
    \Delta h=\max_q h(q)-\min_q h(q),
    \label{eq:mfdfa_width}
\end{equation}
where a larger value indicates greater variation in estimated scaling across moment orders.

The log-moment diagnostic estimates
\begin{equation}
    M_q(\tau)
    =
    \frac{1}{M-\tau}
    \sum_{t=1}^{M-\tau}
    |X_{t+\tau}-X_t|^q
    \propto \tau^{\zeta(q)}.
    \label{eq:log_moment_scaling}
\end{equation}
For each \(q\), \(\zeta(q)\) is obtained from a log-log regression. The moment-scaling function is then approximated by
\begin{equation}
    \zeta(q)=h_2q+cq^2.
    \label{eq:quadratic_zeta}
\end{equation}
The curvature coefficient \(c\), together with the fit of linear and quadratic specifications, summarizes departure from linear moment scaling.

Wavelet-leader analysis uses a Daubechies wavelet and moment orders \(q\in\{-2,-1,1,2,3,4\}\) \citep{wendt2007bootstrap}. An integration of order one is applied before constructing the leaders to satisfy their positive-regularity requirement. This operation shifts the first log-cumulant but leaves the second log-cumulant \(c_2\), the principal wavelet statistic used here, unchanged. A negative \(c_2\) with a 95 percent interval below zero is recorded as evidence of nonlinear wavelet scaling. Full scale choices and bootstrap settings are reported in Appendix A.

\subsubsection{Surrogate Comparisons}
\label{subsubsec:surrogate_comparisons}

The MF-DFA width and absolute log-moment curvature are first compared with shuffled controls in all 32 fixed year-frequency configurations. Shuffling preserves the empirical marginal distribution while removing the recorded temporal order. A detailed comparison for the eight five-minute fixed-window paths additionally uses phase-randomized and iterative amplitude-adjusted Fourier transform surrogates \citep{schreiber2000surrogate,desouza2013components}.

Phase randomization preserves Fourier amplitudes and hence the linear dependence represented by the power spectrum, while generally changing the marginal distribution. Iterative amplitude-adjusted Fourier transform surrogates preserve the empirical marginal distribution and approximate the original power spectrum. For a diagnostic \(D\), the one-sided empirical comparison is
\begin{equation}
    \widehat{p}_{\mathrm{emp}}
    =
    \frac{
        1+
        \displaystyle\sum_{r=1}^{R}
        \mathbf{1}\{D_r^{\mathrm{sur}}\geq D^{\mathrm{obs}}\}
    }{R+1},
    \qquad R=19.
    \label{eq:empirical_p_value}
\end{equation}
With 19 replications, the minimum attainable value is 0.05. These probabilities are interpreted as diagnostic comparisons and considered together across years, statistics, and surrogate types. Because the controls preserve selected features only approximately, the comparisons do not provide a unique causal decomposition of scaling heterogeneity. Algorithmic settings are reported in Appendix A.
\section{Empirical Results}
\label{sec:empirical_results}

\subsection{Pathwise Roughness Across Time and Sampling Frequencies}
\label{subsec:roughness_time_frequency}

Table~\ref{tab:pathwise_roughness_summary} summarizes the principal pathwise estimates. Among the 380 rolling 90-day window-frequency configurations, 341 yield a unique normalized \(p\)-variation root, corresponding to a root rate of 89.7 percent. Root availability is lowest at one minute, where 79 of 95 configurations yield a root, and highest at ten minutes, where 92 of 95 yield a root. The 5- and 15-minute rates lie between these values at 87.4 and 91.6 percent, respectively.

\begin{table}[htbp]
    \centering
    \caption{Pathwise roughness across sampling designs}
    \label{tab:pathwise_roughness_summary}
    \small
    \begin{tabular}{llccc}
        \hline
        Design & Frequency & Unique roots & Root rate, \% & Median \(\widehat{H}\) [IQR] \\
        \hline
        Rolling 90-day & 1 minute  & 79/95 & 83.2 & 0.054 [0.031, 0.073] \\
                       & 5 minutes & 83/95 & 87.4 & 0.065 [0.041, 0.096] \\
                       & 10 minutes & 92/95 & 96.8 & 0.086 [0.062, 0.113] \\
                       & 15 minutes & 87/95 & 91.6 & 0.080 [0.057, 0.118] \\
        \hline
        Non-overlapping 90-day & 1 minute  & 24/32 & 75.0 & 0.073 [0.044, 0.085] \\
                               & 5 minutes & 27/32 & 84.4 & 0.073 [0.043, 0.097] \\
                               & 10 minutes & 31/32 & 96.9 & 0.092 [0.063, 0.112] \\
                               & 15 minutes & 31/32 & 96.9 & 0.076 [0.056, 0.112] \\
        \hline
        Fixed 90-day panel & 1 minute  & 5/8 & 62.5 & 0.065 [0.041, 0.079] \\
                           & 5 minutes & 7/8 & 87.5 & 0.068 [0.032, 0.072] \\
                           & 10 minutes & 7/8 & 87.5 & 0.096 [0.053, 0.102] \\
                           & 15 minutes & 7/8 & 87.5 & 0.097 [0.093, 0.123] \\
        \hline
    \end{tabular}
    \begin{minipage}{0.94\textwidth}
        \footnotesize
        \textit{Notes:} The interquartile range is calculated among configurations with a unique root. Rolling windows begin every 30 days and overlap. The non-overlapping subset retains every third rolling window. The fixed windows form a diagnostic panel and do not represent complete calendar years.
    \end{minipage}
\end{table}

The non-overlapping subset yields 113 unique roots among 128 configurations, or 88.3 percent. Its overall root rate is close to that of the rolling design, although the frequency-specific pattern differs at one and fifteen minutes. Conditional medians also remain low. Across the rolling configurations, median \(\widehat{H}\) ranges from 0.054 at one minute to 0.086 at ten minutes. The corresponding non-overlapping medians range from 0.073 to 0.092.

The fixed diagnostic panel provides supporting evidence from a smaller set of common yearly windows. A unique root is found in 26 of 32 configurations. The remaining six configurations have no root on the specified search interval, and no configuration has multiple roots. The no-root cases occur three times at one minute and once at each aggregated frequency. All finite estimates in Table~\ref{tab:pathwise_roughness_summary} are below \(1/2\). More broadly, every finite estimate obtained in the rolling, non-overlapping, alternative-length, and jump-robust analyses is below \(1/2\), with the largest equal to approximately 0.193.

\begin{figure}[htbp]
    \centering
    \begin{minipage}{0.49\textwidth}
        \centering
        \includegraphics[width=\textwidth]{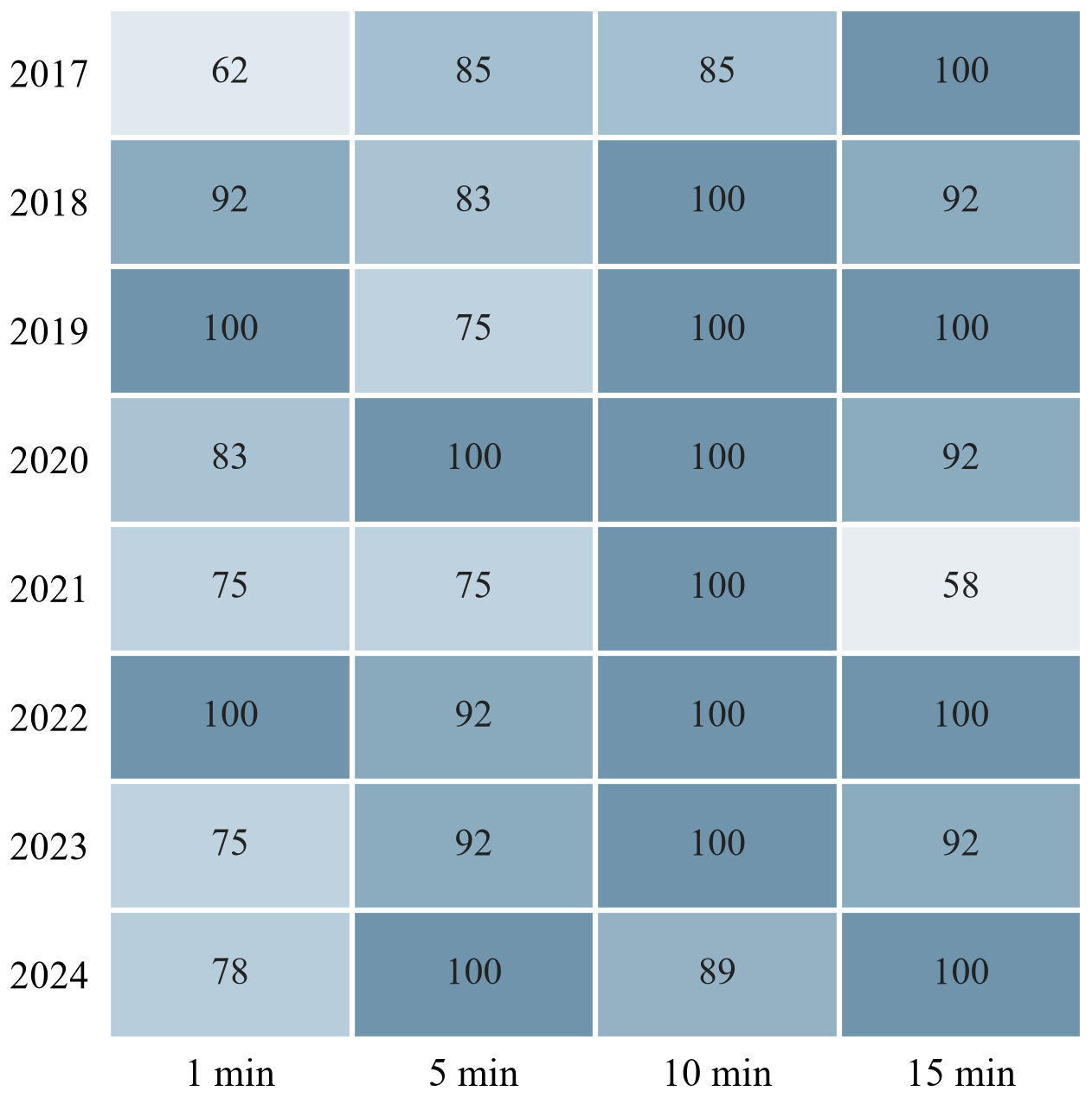}

        \small (a) Rolling unique-root rates
    \end{minipage}
    \begin{minipage}{0.49\textwidth}
        \centering
        \includegraphics[width=\textwidth]{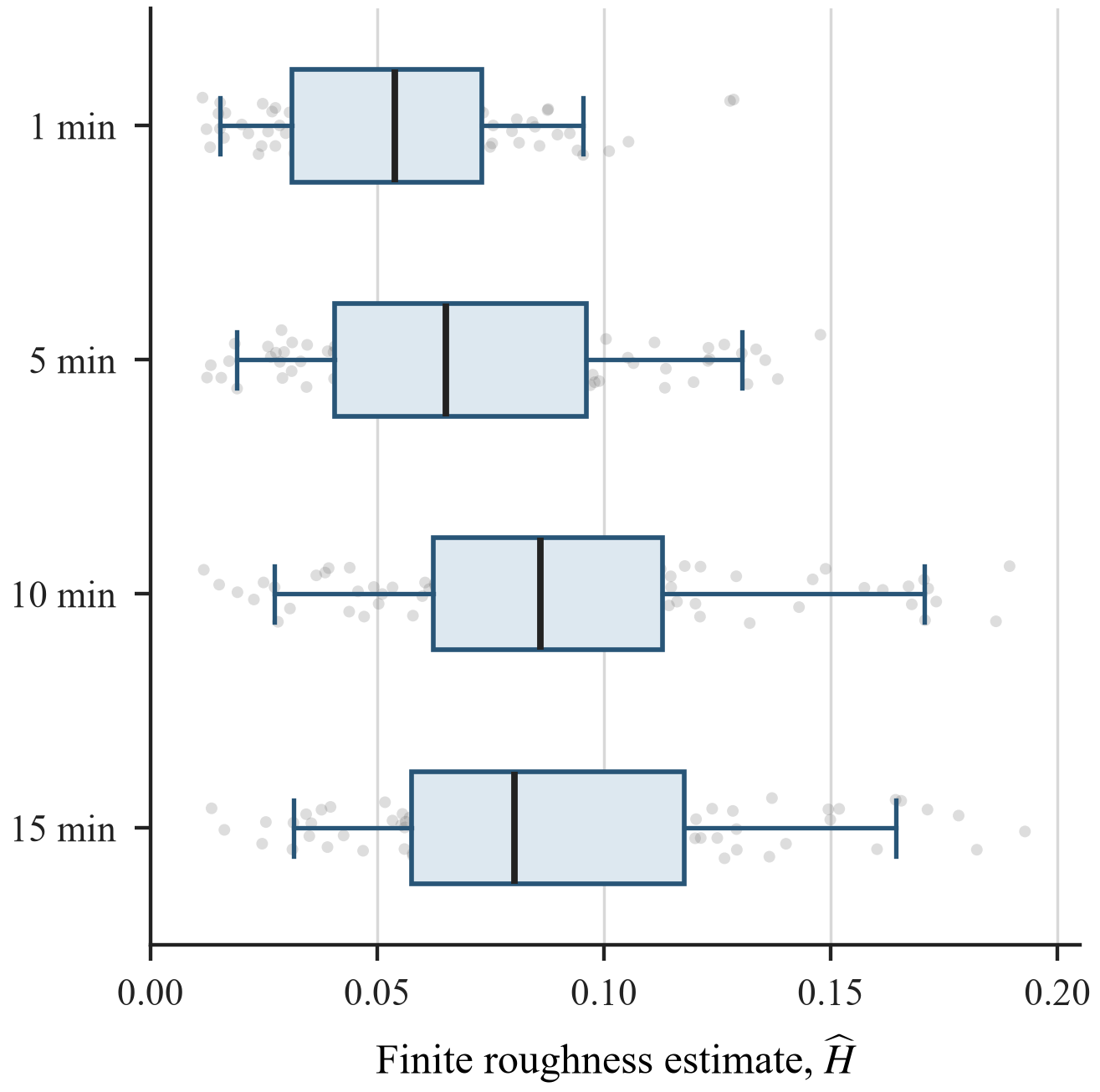}

        \small (b) Finite roughness estimates
    \end{minipage}
    \caption{Root availability and finite \(\widehat{H}\) in rolling 90-day windows}
    \label{fig:rolling_roughness}
\end{figure}

Figure~\ref{fig:rolling_roughness} shows that the aggregate root rates conceal variation across individual windows. The finite estimates are concentrated well below \(1/2\), but their distributions overlap only partially across frequencies and contain substantial within-frequency dispersion. Root existence is therefore common in the observed samples, while neither root availability nor the value of \(\widehat{H}\) is constant across time and sampling frequency.

\subsection{Sensitivity to Sample Horizon and Volatility Construction}
\label{subsec:sensitivity_horizon_measure}

The fixed-window outcomes do not consistently occupy the center of their same-year rolling distributions. Fourteen of the 32 fixed year-frequency configurations yield a finite \(\widehat{H}\) within the interquartile range of rolling estimates whose windows begin in the same year. Twelve yield a finite estimate outside that range, and six have no root. The fixed diagnostic panel therefore complements the broader rolling evidence by holding dates constant across diagnostic methods.

Panel A of Table~\ref{tab:measurement_sensitivity} reports the window-length comparison. Unique roots are obtained in 28 of 32 configurations at 60 days, 26 of 32 at 90 days, and 29 of 32 at 120 days. Root availability remains high at each horizon, but it does not change monotonically with window length. Conditional medians also vary. At one minute, for example, the median is 0.033 at 60 days, 0.065 at 90 days, and 0.028 at 120 days. At fifteen minutes, the corresponding medians are 0.077, 0.097, and 0.062. These differences show that extending the path can change the estimated variation order even when a unique root remains available.

\begin{table}[p]
    \centering
    \caption{Sensitivity to window length and volatility construction}
    \label{tab:measurement_sensitivity}
    \small
    \textit{Panel A: Window length} \\[3pt]
    \begin{tabular}{lccc}
        \hline
        Frequency & 60 days & 90 days & 120 days \\
        \hline
        1 minute  & 6/8; 0.033 & 5/8; 0.065 & 7/8; 0.028 \\
        5 minutes & 7/8; 0.080 & 7/8; 0.068 & 8/8; 0.056 \\
        10 minutes & 7/8; 0.088 & 7/8; 0.096 & 7/8; 0.069 \\
        15 minutes & 8/8; 0.077 & 7/8; 0.097 & 7/8; 0.062 \\
        \hline
    \end{tabular}

    \vspace{10pt}
    \textit{Panel B: Volatility construction in fixed 90-day windows} \\[3pt]
    \begin{tabular}{lcccc}
        \hline
        Measure & 1 minute & 5 minutes & 10 minutes & 15 minutes \\
        \hline
        Baseline RV & 5/8; 0.065 & 7/8; 0.068 & 7/8; 0.096 & 7/8; 0.097 \\
        Truncated RV, \(4\widehat{\sigma}\) & 4/8; 0.020 & 6/8; 0.036 & 7/8; 0.061 & 7/8; 0.094 \\
        Truncated RV, \(6\widehat{\sigma}\) & 4/8; 0.035 & 7/8; 0.050 & 7/8; 0.070 & 7/8; 0.092 \\
        Truncated RV, \(8\widehat{\sigma}\) & 3/8; 0.043 & 6/8; 0.035 & 7/8; 0.047 & 7/8; 0.093 \\
        Bipower variation & Not applicable & 5/8; 0.019 & 7/8; 0.088 & 8/8; 0.112 \\
        \hline
    \end{tabular}
    \begin{minipage}{0.94\textwidth}
        \footnotesize
        \textit{Notes:} Each entry reports unique roots over configurations, followed by median \(\widehat{H}\) among finite estimates. Panel A holds the eight starting dates fixed. Panel B uses the eight fixed 90-day windows. RV denotes realized volatility.
    \end{minipage}
\end{table}

Panel B examines sensitivity to extreme-return treatment. At the intermediate truncation threshold of six local-scale units, flagged returns account for between 0.27 and 1.88 percent of observations across the eight fixed windows, but for between 10.1 and 32.8 percent of squared variation. Truncation therefore modifies a limited share of returns with a comparatively large contribution to realized variation.

The effect on root availability is most pronounced at one minute. The baseline rate of 62.5 percent falls to 50.0 percent at the four- and six-unit thresholds and to 37.5 percent at the eight-unit threshold. At 5, 10, and 15 minutes, roots remain available in most configurations under every truncation rule. The ordering across thresholds is not monotonic, which indicates that root existence depends on the resulting path rather than only on how many returns are removed.

\begin{figure}[htbp]
    \centering
    \includegraphics[width=0.88\textwidth]{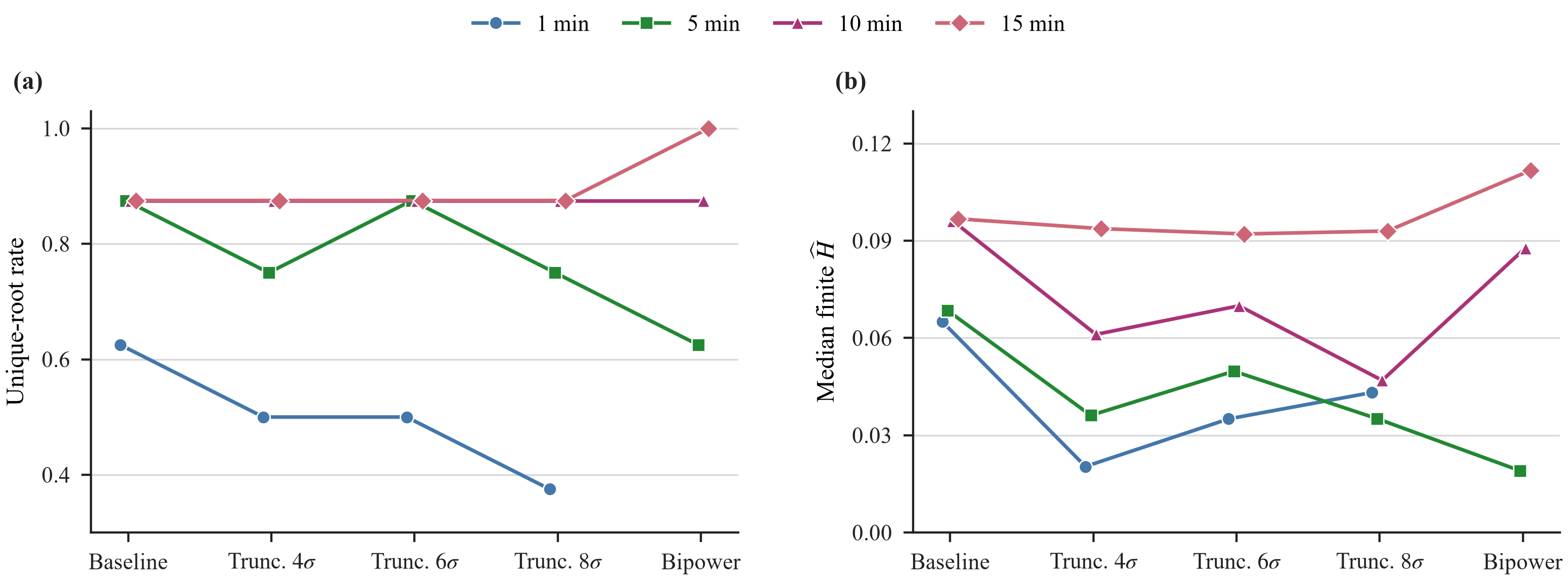}
    \caption{Root availability under baseline and jump-robust volatility measures}
    \label{fig:jump_robustness}
\end{figure}

Bipower variation yields a unique root in 20 of the 24 eligible configurations, compared with 21 of 24 for baseline realized volatility at the same frequencies. The median estimate falls from 0.068 to 0.019 at five minutes, changes from 0.096 to 0.088 at ten minutes, and rises from 0.097 to 0.112 at fifteen minutes. Jump-robust constructions thus retain low finite estimates at aggregated intraday frequencies, but they can materially alter their magnitude. Figure~\ref{fig:jump_robustness} summarizes these comparisons.

\subsection{Temporal Evolution and Structural Variation}
\label{subsec:temporal_evolution_structural}

\begin{figure}[htbp]
    \centering
    \includegraphics[width=0.90\textwidth]{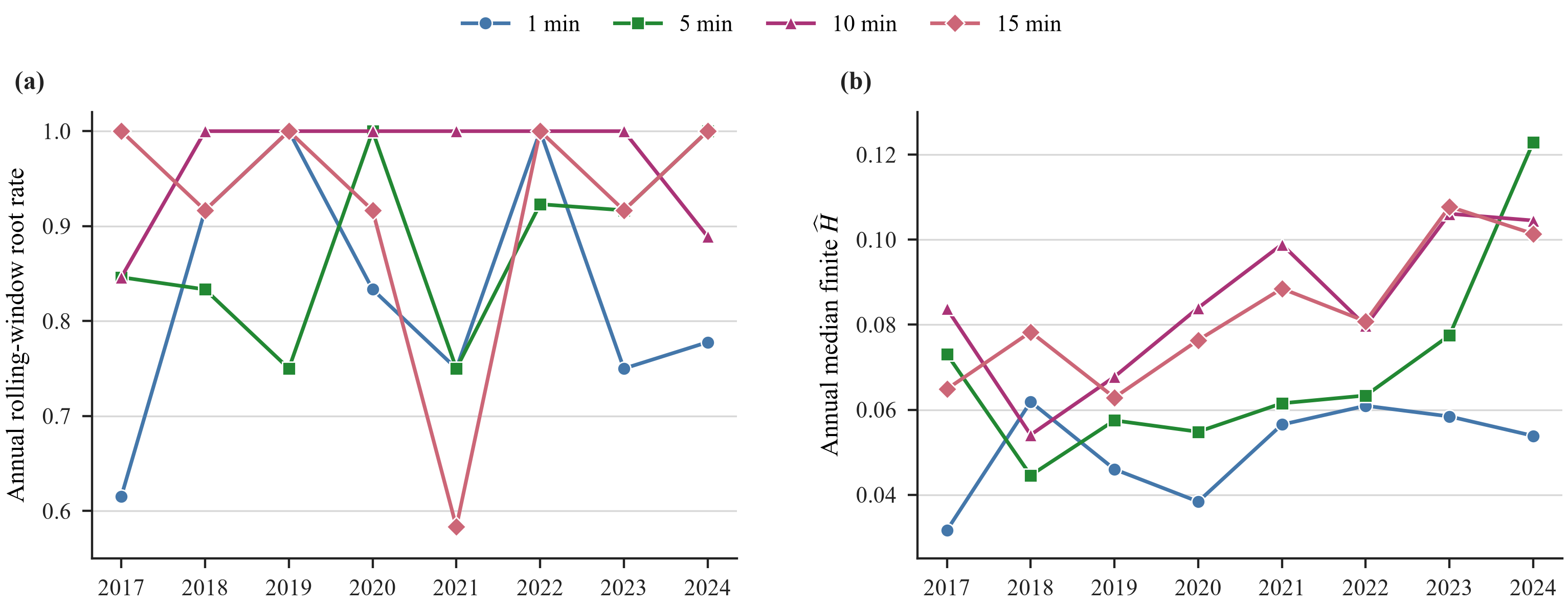}
    \caption{Temporal evolution of rolling root availability and finite roughness estimates}
    \label{fig:roughness_temporal_evolution}
\end{figure}

Annual summaries of the rolling windows show no significant monotonic trend in unique-root availability at any frequency. Median finite \(\widehat{H}\) has a positive Spearman correlation with time at ten minutes, with \(\rho=0.714\) and \(p=0.047\), and at fifteen minutes, with \(\rho=0.833\) and \(p=0.010\). The associated Theil-Sen point estimates are increases of 0.0058 and 0.0059 per year, respectively. The one- and five-minute median estimates do not exhibit significant monotonic trends. Given the eight annual summaries, these results indicate selected frequency-specific changes rather than a common temporal trend. These annual summaries are shown in Figure~\ref{fig:roughness_temporal_evolution}.

\begin{figure}[htbp]
    \centering
    \includegraphics[width=0.92\textwidth]{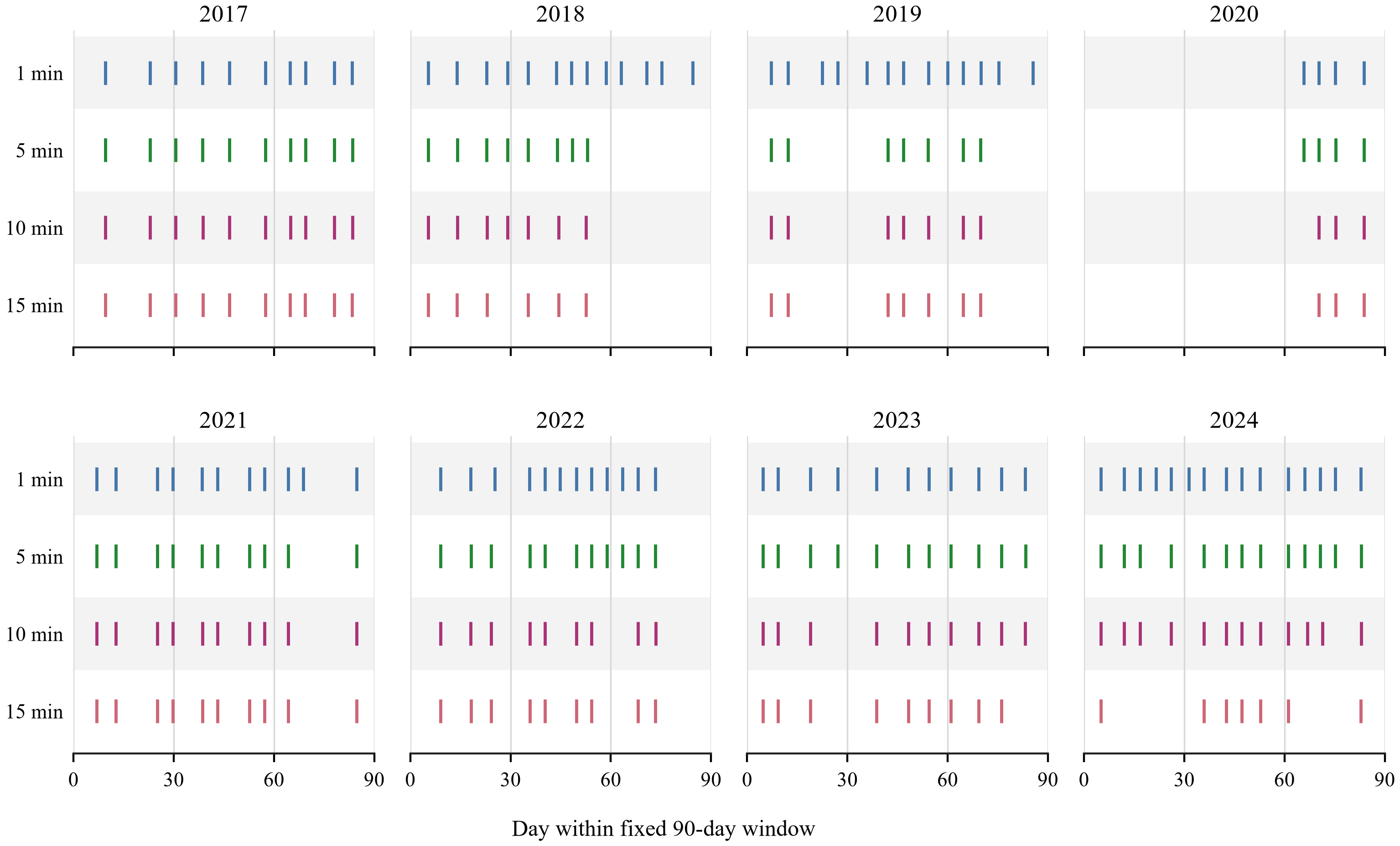}
    \caption{Penalized structural-break map for fixed realized-volatility paths}
    \label{fig:break_map}
\end{figure}

The fixed-window volatility levels display frequent structural variation. Penalized binary segmentation identifies 292 breakpoints across the 32 configurations, with a median of 10 per configuration and a range from 3 to 15. Figure~\ref{fig:break_map} shows that detected changes occur across years and frequencies rather than being confined to a single period or resolution. Since frequency-specific paths within each window derive from the same return base, the break counts are descriptive and are not treated as independent events.

The supporting level tests provide a similar indication that a constant-level characterization is inadequate for these fixed samples. The augmented Dickey-Fuller test rejects its unit-root null in all 32 configurations, while the Kwiatkowski-Phillips-Schmidt-Shin test also rejects its level-stationarity null in all 32. No configuration therefore meets the joint criterion defined in Section~\ref{subsubsec:volatility_level_breaks}. These outcomes are compatible with substantial level shifts or other departures from either test's maintained specification.

\subsection{Scaling Heterogeneity and Surrogate Evidence}
\label{subsec:scaling_surrogate_results}

The three scaling diagnostics vary across the fixed year-frequency panel. Relative to shuffled controls, MF-DFA width reaches the minimum empirical probability of 0.05 in 18 of 32 configurations, while absolute log-moment curvature does so in 25 of 32. The 95 percent interval for the wavelet second log-cumulant lies below zero in 14 configurations. All three conditions occur together in 12 configurations. These counts indicate heterogeneous scaling in several samples, but the shuffled comparison controls only for the empirical marginal distribution.

The detailed five-minute analysis provides a more restrictive comparison. Table~\ref{tab:surrogate_counts} reports the number of yearly samples in which the observed statistic exceeds all 19 surrogate statistics. Against shuffled surrogates, excess MF-DFA width occurs in four of eight samples and excess absolute log-moment curvature in five. Against phase-randomized controls, each statistic is significant in seven samples. Relative to IAAFT controls, which retain the empirical marginal distribution and approximate the original power spectrum jointly, the counts fall to three for MF-DFA width and one for log-moment curvature.

\begin{table}[htbp]
    \centering
    \caption{Five-minute scaling diagnostics relative to surrogate controls}
    \label{tab:surrogate_counts}
    \small
    \begin{tabular}{lcc}
        \hline
        Surrogate control & Excess MF-DFA width & Excess log-moment curvature \\
        \hline
        Shuffled & 4/8 & 5/8 \\
        Phase-randomized & 7/8 & 7/8 \\
        IAAFT & 3/8 & 1/8 \\
        \hline
    \end{tabular}
    \begin{minipage}{0.90\textwidth}
        \footnotesize
        \textit{Notes:} Counts use the one-sided empirical probability in Equation~\eqref{eq:empirical_p_value} and the 0.05 threshold. With 19 replications, 0.05 is the minimum attainable value. IAAFT denotes iterative amplitude-adjusted Fourier transform.
    \end{minipage}
\end{table}

\begin{figure}[htbp]
    \centering
    \includegraphics[width=0.92\textwidth]{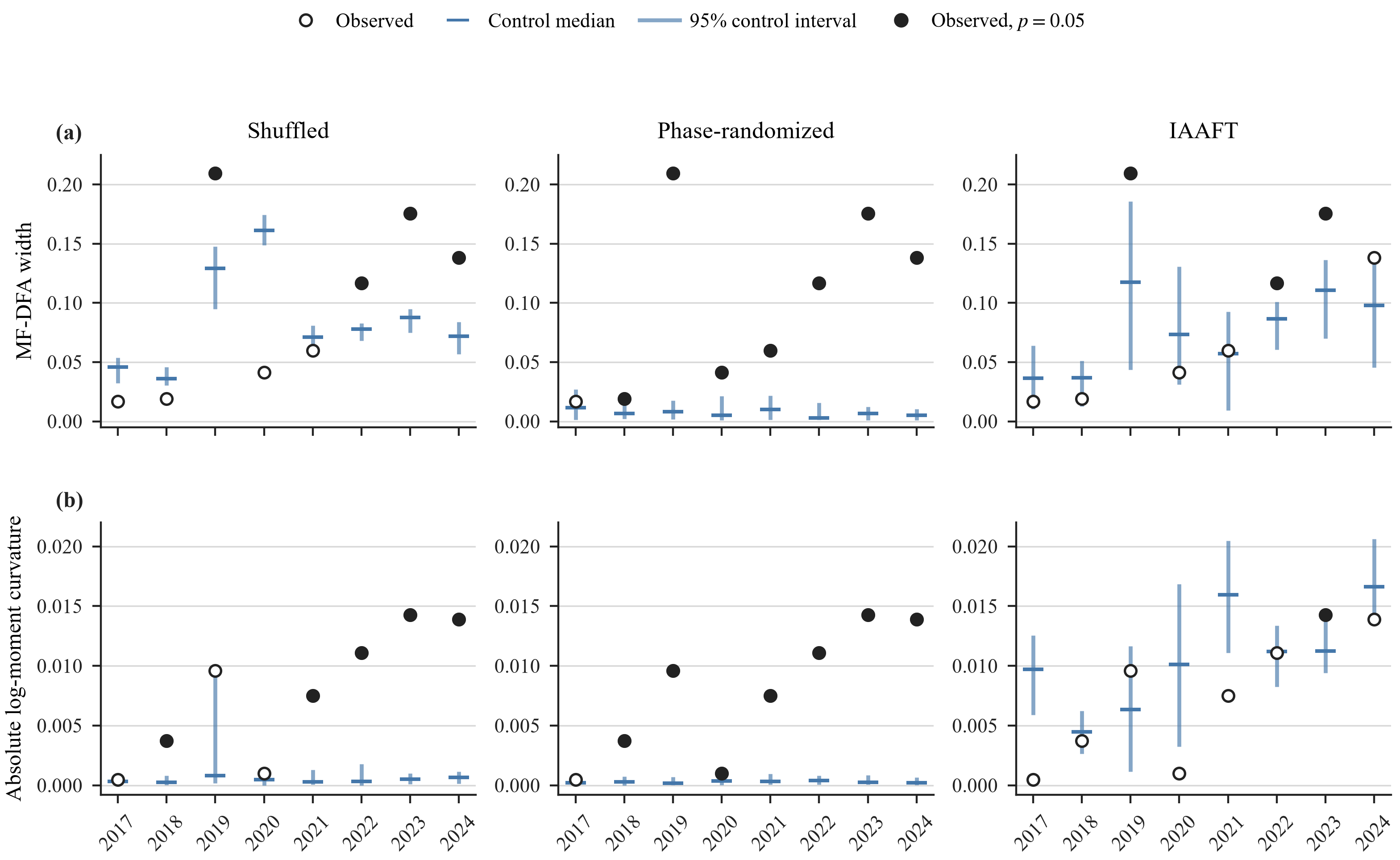}
    \caption{Observed five-minute scaling diagnostics and surrogate distributions}
    \label{fig:surrogate_comparison}
\end{figure}

Figure~\ref{fig:surrogate_comparison} displays the observed statistics and the corresponding surrogate distributions. The contrast across surrogate types indicates that the marginal distribution and linear dependence reproduce a substantial portion of the observed scaling statistics. Evidence beyond both features is present in selected years, but it is not uniform across the five-minute panel. The comparison does not assign the residual to a unique nonlinear mechanism, particularly given the finite number of replications and the approximate spectral match of IAAFT controls.

Temporal patterns in the raw diagnostics are also mixed. MF-DFA width and wavelet \(c_2\) show no significant monotonic trend at any frequency. Signed log-moment curvature becomes more negative at all four frequencies, with each Spearman test reaching the 5 percent threshold. Because the three diagnostics do not move uniformly, the results do not indicate a general decline in scaling heterogeneity between 2017 and 2024.

\section{Discussion}
\label{sec:discussion}

\subsection{Interpretation of the Main Evidence}
\label{subsec:interpretation_main_evidence}

The principal result is that Bitcoin realized volatility usually admits a unique normalized \(p\)-variation root. Root existence is common in both the rolling design and the non-overlapping subset, and all finite estimates lie below \(1/2\). The evidence therefore supports low pathwise regularity in the observed realized-volatility series. At the same time, the absence of a root in some configurations shows that a finite estimate is not available for every period and measurement design.

Although root availability is high overall, the finite estimates are not constant. Their distributions vary across windows and frequencies, and fixed-window estimates frequently lie outside the interquartile range of same-year rolling estimates. An estimate from one sample may consequently differ from the range observed in nearby periods.

The frequency pattern is also relevant. Rolling median \(\widehat{H}\) is lowest at one minute and higher at the aggregated frequencies, while root availability is lowest at one minute and highest at ten minutes. Aggregation changes the realized-volatility path, reduces the number of observations, and alters the relative influence of individual returns. The observed frequency differences cannot be assigned to any one of these channels, but they indicate that estimated roughness is partly a property of the measurement scale.

The jump-robust comparisons further qualify the estimates. Locally truncated realized volatility changes both root availability and finite \(\widehat{H}\), with the largest reduction in root availability occurring at one minute. Bipower variation preserves roots in most eligible configurations but changes the median estimate, particularly at five minutes. Large returns therefore coincide with meaningful changes in measured path regularity. The comparisons do not isolate a causal jump effect because truncation and bipower variation also change the construction and local dependence of the volatility path.

Temporal diagnostics provide additional context. Root availability has no common monotonic trend, while median finite \(\widehat{H}\) increases at the two coarser frequencies only. The structural-break analysis identifies repeated shifts in realized-volatility levels throughout the sample. These results are consistent with a market whose local volatility conditions vary over time, but they do not imply that the variation index and the volatility level follow the same dynamics. In particular, the level diagnostics are descriptive complements to the pathwise estimator rather than conditions for its use.

The scaling diagnostics do not provide uniform evidence of residual heterogeneity after controlling for the marginal distribution and linear dependence. Raw diagnostics and shuffled comparisons identify scaling heterogeneity in several configurations, but the detailed five-minute results change substantially when both the marginal distribution and approximate linear dependence are preserved. Residual MF-DFA width remains in three yearly samples and residual log-moment curvature in one. The features reproduced by the IAAFT controls therefore account for much of the observed statistics, while additional scaling structure remains possible in selected periods. This conclusion is conditional on the diagnostics, controls, and finite samples considered here.

\subsection{Modeling and Practical Implications}
\label{subsec:modeling_practical_implications}

The results favor treating the roughness index as an empirical feature whose stability should be assessed rather than presumed. A constant-\(H\) specification may provide an effective approximation in a given sample, but the variation across windows, frequencies, and volatility measures suggests that its calibration can depend on the observation design. Models with time-varying roughness or regime-sensitive parameters offer one possible response to temporal variation. Rough volatility models with jumps provide another way to represent short-scale irregularity together with discontinuous returns. Hybrid rough and multiscale specifications may be useful when a single scaling relation does not summarize all moments and horizons.

These model classes are directions for comparison rather than implications established by the pathwise estimates alone. Their empirical relevance should be assessed directly through likelihood-based fit, out-of-sample volatility forecasts, option-pricing errors, or risk-measure performance. Such comparisons would determine whether allowing for jumps, regimes, or multiple scaling components improves economically relevant outcomes relative to a simpler rough-volatility benchmark.

The measurement sensitivity also has practical relevance. Roughness estimates may enter volatility forecasting, derivative valuation, hedging, and the calibration of risk models for Bitcoin-linked positions. If the estimate changes with sampling frequency, sample horizon, or jump treatment, downstream model outputs may change as well. Empirical applications can address this issue by reporting the volatility construction and estimation window explicitly and by examining whether conclusions persist across reasonable alternatives. The present analysis documents variation in the input estimate; it does not quantify the resulting differences in prices, hedge ratios, or risk forecasts.

\subsection{Limitations and Future Research}
\label{subsec:limitations_future_research}

The first limitation concerns data scope. The analysis uses BTC/USD close prices from one exchange and a one-minute base frequency. Other venues differ in liquidity, trading intensity, and price formation, while quote, trade, or order-book data could support alternative treatments of market microstructure. The volatility paths are realized measures derived from recorded returns and do not reveal latent spot volatility directly. Future work could compare venues, use finer transaction-level data, and combine pathwise diagnostics with estimators designed explicitly for latent volatility and observation error.

The second limitation concerns the sampling design. Rolling windows provide broad temporal coverage but overlap, so their configuration counts are descriptive rather than independent-sample frequencies. The non-overlapping subset reduces this dependence but contains fewer periods. The fixed yearly windows support common-date diagnostics but do not represent complete calendar years. Temporal trend tests use eight annual summaries, and the detailed surrogate analysis uses eight five-minute samples with 19 replications per control. A longer observation period would provide more non-overlapping windows for temporal comparison, while larger surrogate ensembles would permit more precise surrogate comparisons. Formal adjustment for the related set of diagnostic comparisons would also be useful.

The third limitation concerns method scope. The baseline partition follows a single sample-size rule, and the partition-sensitivity exercise covers one fixed 2021 window. Broader analysis across partition sequences would clarify how often root availability and finite estimates change with the coarse-block design. The stated search interval also conditions the classification of root availability. Finally, the study does not compare structural volatility models through forecasting, pricing, or formal model selection. Future research can use the documented pathwise estimates and temporal variation to formulate candidate rough, jump, regime-sensitive, and multiscale models, then evaluate them against common economic loss functions.
\section{Conclusion}
\label{sec:conclusion}

This paper examines whether Bitcoin realized volatility admits a measurable pathwise roughness index and whether the estimate is stable across time and measurement designs. The analysis applies the normalized \(p\)-variation estimator to realized-volatility paths constructed from one-minute BTC/USD close prices between 2017 and 2024.

A unique root is obtained in 89.7 percent of the rolling 90-day window-frequency configurations and 88.3 percent of the non-overlapping subset. All finite estimates from the temporal, window-length, and jump-robust analyses are below \(1/2\). Bitcoin realized volatility therefore generally admits a low pathwise roughness estimate in the configurations examined. Root availability is not universal, however, and the magnitude of finite \(\widehat{H}\) varies across periods, sampling frequencies, and window lengths.

The volatility construction also matters. Truncating extreme returns reduces root availability most clearly at one minute and changes finite estimates across frequencies. Bipower variation retains a unique root in most eligible fixed-window configurations, with low estimates continuing to appear at 5-, 10-, and 15-minute frequencies. Temporal diagnostics identify substantial variation in realized-volatility levels, while root availability has no significant monotonic trend at any frequency.

Scaling diagnostics qualify the interpretation of the pathwise estimates. In the five-minute fixed-window panel, IAAFT controls reproduce much of the observed scaling heterogeneity, while the remaining differences are period-specific and do not identify a unique source of nonlinear scaling.

The central conclusion is that Bitcoin realized volatility usually exhibits low pathwise roughness, but the availability and magnitude of the estimate depend on the observed period and measurement procedure. These findings concern realized volatility constructed from discrete returns and do not directly identify the roughness of latent spot volatility.

\appendix

\appendix

\section{Methodological and Numerical Details}
\label{app:methodological_details}

\subsection{Normalized \texorpdfstring{\(p\)}{p}-Variation Algorithm}
\label{app:roughness_algorithm}

For each realized-volatility path, the implementation proceeds as follows:

\begin{enumerate}
    \item Retain the longest contiguous sequence of finite volatility observations and denote its number of fine increments by \(N\).
    \item Set the baseline number of coarse blocks to
    \[
        K=\max\left\{2,\left\lfloor\sqrt{N}\right\rfloor\right\},
    \]
    define \(n=\lfloor N/K\rfloor\), and use the first \(L=Kn\) increments. Observations in the unused terminal remainder are excluded.
    \item Normalize the observation interval to \([0,1]\) and evaluate Equation~\eqref{eq:normalized_p_variation} over \(H\in[0.01,0.95]\), using \(p=1/H\).
    \item Compute \(\log W(L,K,p)\) with log-sum-exp operations. This avoids direct exponentiation of high powers and reduces numerical overflow and underflow.
    \item Identify sign changes in \(\log W\) on the initial grid and refine each crossing with Brent's bracketing method.
    \item Report \(\widehat{H}\) only when exactly one root is detected. Record configurations with zero or multiple roots separately.
\end{enumerate}

Table~\ref{tab:numerical_settings} summarizes the principal settings.

\begin{table}[htbp]
    \centering
    \caption{Principal numerical settings}
    \label{tab:numerical_settings}
    \small
    \begin{tabular}{p{0.35\textwidth}p{0.55\textwidth}}
        \hline
        Component & Setting \\
        \hline
        Admissible roughness interval & \(H\in[0.01,0.95]\), implying \(p=1/H>1\) \\
        Baseline coarse blocks & \(K=\max\{2,\lfloor\sqrt{N}\rfloor\}\) \\
        Fixed-panel initial grid & 320 equally spaced \(H\) values \\
        Rolling and window-length grid & 160 equally spaced \(H\) values \\
        Root refinement & Brent's bracketing method \\
        Numerical evaluation & Logarithmic form with log-sum-exp aggregation \\
        Root classification & Unique root, no root, or multiple roots \\
        \hline
    \end{tabular}
\end{table}

\subsection{Fractional Brownian Motion Implementation Check}
\label{app:fbm_check}

The numerical implementation is evaluated on fractional Brownian motion paths generated by the Davies-Harte method \citep{davies1987tests}. Each path contains 16,384 increments. Six independent paths are generated for each true Hurst parameter \(H_0\in\{0.10,0.30,0.50\}\), and the baseline partition and root procedure used for the fixed empirical panel are applied. Table~\ref{tab:fbm_check} reports the resulting estimates.

\begin{table}[htbp]
    \centering
    \caption{Fractional Brownian motion implementation check}
    \label{tab:fbm_check}
    \small
    \begin{tabular}{cccccc}
        \hline
        \(H_0\) & Valid roots & Mean \(\widehat{H}\) & SD & Mean error & Maximum absolute error \\
        \hline
        0.10 & 6/6 & 0.1233 & 0.0196 & 0.0233 & 0.0464 \\
        0.30 & 6/6 & 0.3004 & 0.0167 & 0.0004 & 0.0276 \\
        0.50 & 6/6 & 0.5075 & 0.0104 & 0.0075 & 0.0214 \\
        \hline
    \end{tabular}
\end{table}

\subsection{Partition Sensitivity in the Fixed 2021 Window}
\label{app:k_sensitivity}

The limited partition check multiplies the baseline \(K\) by \(0.60\), \(0.75\), \(0.90\), \(1.00\), \(1.10\), \(1.25\), and \(1.40\), with the result rounded to the nearest integer. Table~\ref{tab:k_sensitivity} reports the resulting estimate and the associated value of \(K\).

\begin{table}[htbp]
    \centering
    \caption{Partition sensitivity for the fixed 2021 window}
    \label{tab:k_sensitivity}
    \scriptsize
    \setlength{\tabcolsep}{3.5pt}
    \begin{tabular}{lccccccc}
        \hline
        Frequency & \(0.60K_0\) & \(0.75K_0\) & \(0.90K_0\) & \(K_0\) & \(1.10K_0\) & \(1.25K_0\) & \(1.40K_0\) \\
        \hline
        1 minute &
        0.0267 (215) & 0.0360 (269) & 0.0260 (323) & 0.0838 (359) &
        0.0269 (395) & NR (449) & NR (503) \\
        5 minutes &
        0.0460 (96) & NR (120) & 0.1264 (144) & 0.0129 (160) &
        0.2025 (176) & 0.0479 (200) & 0.0846 (224) \\
        10 minutes &
        0.0825 (68) & 0.0909 (85) & 0.1253 (102) & 0.1029 (113) &
        0.1160 (124) & 0.0533 (141) & 0.0965 (158) \\
        15 minutes &
        0.0879 (55) & 0.0116 (69) & NR (83) & 0.0984 (92) &
        0.0836 (101) & 0.0921 (115) & 0.0972 (129) \\
        \hline
    \end{tabular}

    \begin{minipage}{0.94\textwidth}
        \footnotesize
        \textit{Notes:} Entries report \(\widehat{H}\), followed by \(K\) in parentheses. \(K_0\) is the baseline value. NR denotes no root on the stated search interval.
    \end{minipage}
\end{table}

\subsection{Jump-Robustness Settings}
\label{app:jump_settings}

The local truncation scale uses the median absolute one-minute return over the preceding 1,440 minutes, divided by \(\Phi^{-1}(0.75)\). The rolling median is shifted by one minute so that the current return does not enter its own threshold. At least 720 prior observations are required. When this history is unavailable at the beginning of an analysis window, the median-based scale calculated over the complete window is used. Returns satisfying
\[
    |r_t|>c\widehat{\sigma}_t,
    \qquad c\in\{4,6,8\},
\]
are set to zero before the remaining squared returns are aggregated. Each \(m\)-minute volatility observation is retained only when all \(m\) constituent returns are available.

For bipower variation, adjacent absolute-return products are formed only within the same aggregation bin. The first return in each bin has no preceding within-bin return and is excluded from the product sum. A complete \(m\)-minute bipower observation therefore contains \(m-1\) adjacent products.

\subsection{Temporal and Structural Settings}
\label{app:temporal_settings}

The augmented Dickey-Fuller regressions include a constant. For a volatility path with \(M\) observations, the maximum candidate lag is
\[
    \min\left\{
        20,
        \max\left(1,\left\lfloor M/100\right\rfloor\right)
    \right\},
\]
and the selected lag minimizes the Akaike information criterion. The Kwiatkowski-Phillips-Schmidt-Shin regressions also include a constant and use automatic lag selection.

Penalized binary segmentation is applied after standardizing each path. The cost is the within-segment sum of squared deviations. The minimum segment length is
\[
    \max\left\{20,\left\lfloor0.05M\right\rfloor\right\},
\]
candidate split locations are evaluated every
\[
    \max\left\{1,\left\lfloor M/5000\right\rfloor\right\}
\]
observations, and the penalty is \(3\log M\). Break locations are converted back to UTC timestamps after estimation.

Annual temporal patterns are evaluated with Theil-Sen slopes and two-sided Spearman rank correlations. The roughness trends use annual summaries of the rolling windows. Scaling trends use the eight fixed-window values at each frequency.

\subsection{Scaling and Surrogate Settings}
\label{app:scaling_settings}

MF-DFA and log-moment scaling use \(q\in\{0.5,1.0,1.5,2.0,2.5,3.0\}\). Fourteen unique integer scales are selected on a logarithmic grid. For sampling interval \(m\) minutes and path length \(M\), the lower and upper bounds are
\[
    s_{\min}=\max\left\{8,\left\lceil60/m\right\rceil\right\},
    \qquad
    s_{\max}=
    \min\left\{
        \left\lfloor2880/m\right\rfloor,
        \left\lfloor M/16\right\rfloor
    \right\}.
\]
MF-DFA uses first-order polynomial detrending. Log-moment scaling uses the same physical scale range and reports both the linear fit and the quadratic coefficient \(c\) in Equation~\eqref{eq:quadratic_zeta}.

Wavelet leaders use a Daubechies wavelet with three vanishing moments, \(q\in\{-2,-1,1,2,3,4\}\), and an integration of order one before leader construction. Regressions are weighted by the number of available coefficients at each dyadic scale. The reported 95 percent interval for \(c_2\) uses the standard deviation from 39 bootstrap replications and a normal approximation.

Each surrogate comparison uses 19 replications. Shuffled controls randomly permute the observed values. Phase-randomized controls retain Fourier amplitudes and randomize the nontrivial phases. IAAFT controls iteratively restore the empirical ranks and Fourier amplitudes. The IAAFT procedure uses at most 200 iterations and stops earlier if the rank ordering no longer changes or if the improvement in relative spectral error is no greater than \(10^{-7}\). The detailed three-surrogate comparison is restricted to the eight five-minute fixed-window paths.

\section{Supplementary Empirical Results}
\label{app:supplementary_results}

\subsection{Fixed Diagnostic Windows}
\label{app:data_audit}

\begin{table}[htbp]
    \centering
    \caption{Fixed 90-day diagnostic windows}
    \label{tab:fixed_windows_appendix}
    \small
    \begin{tabular}{ccc}
        \hline
        Year & Start, UTC & End, UTC \\
        \hline
        2017 & 2017-10-03 00:00 & 2017-12-31 23:59 \\
        2018 & 2018-02-13 00:00 & 2018-05-13 23:59 \\
        2019 & 2019-05-11 00:00 & 2019-08-08 23:59 \\
        2020 & 2020-01-02 00:00 & 2020-03-31 23:59 \\
        2021 & 2021-01-01 00:00 & 2021-03-31 23:59 \\
        2022 & 2022-01-03 00:00 & 2022-04-02 23:59 \\
        2023 & 2023-01-08 00:00 & 2023-04-07 23:59 \\
        2024 & 2024-01-22 00:00 & 2024-04-20 23:59 \\
        \hline
    \end{tabular}
\end{table}

Each fixed window contains 129,600 valid one-minute returns. It produces 129,600, 25,920, 12,960, and 8,640 complete realized-volatility observations at the 1-, 5-, 10-, and 15-minute frequencies, respectively. Exact zeros remain in the volatility paths and are not removed.

\clearpage

\subsection{Fixed-Window Pathwise Estimates}
\label{app:fixed_roughness}

Table~\ref{tab:fixed_roughness_appendix} reports all 32 baseline estimates in the fixed diagnostic panel. A unique root is obtained in 26 configurations. Six configurations have no root, and none has multiple roots.

\begingroup
\small
\begin{longtable}{cclcc}
    \caption{Complete fixed-window normalized \(p\)-variation results}
    \label{tab:fixed_roughness_appendix}\\
    \hline
    Year & Frequency & Root status & \(\widehat{H}\) & \(\widehat{p}\) \\
    \hline
    \endfirsthead
    \multicolumn{5}{c}{\tablename\ \thetable\ continued} \\
    \hline
    Year & Frequency & Root status & \(\widehat{H}\) & \(\widehat{p}\) \\
    \hline
    \endhead
    2017 & 1 minute  & No root & & \\
    2017 & 5 minutes & Unique & 0.0369 & 27.104 \\
    2017 & 10 minutes & No root & & \\
    2017 & 15 minutes & No root & & \\
    2018 & 1 minute  & Unique & 0.0787 & 12.709 \\
    2018 & 5 minutes & No root & & \\
    2018 & 10 minutes & Unique & 0.0183 & 54.522 \\
    2018 & 15 minutes & Unique & 0.0925 & 10.815 \\
    2019 & 1 minute  & Unique & 0.0650 & 15.376 \\
    2019 & 5 minutes & Unique & 0.0685 & 14.607 \\
    2019 & 10 minutes & Unique & 0.1111 & 9.001 \\
    2019 & 15 minutes & Unique & 0.1473 & 6.787 \\
    2020 & 1 minute  & No root & & \\
    2020 & 5 minutes & Unique & 0.0729 & 13.713 \\
    2020 & 10 minutes & Unique & 0.0491 & 20.378 \\
    2020 & 15 minutes & Unique & 0.0541 & 18.477 \\
    2021 & 1 minute  & Unique & 0.0838 & 11.940 \\
    2021 & 5 minutes & Unique & 0.0129 & 77.273 \\
    2021 & 10 minutes & Unique & 0.1029 & 9.715 \\
    2021 & 15 minutes & Unique & 0.0984 & 10.163 \\
    2022 & 1 minute  & Unique & 0.0384 & 26.023 \\
    2022 & 5 minutes & Unique & 0.0281 & 35.610 \\
    2022 & 10 minutes & Unique & 0.0572 & 17.491 \\
    2022 & 15 minutes & Unique & 0.0968 & 10.335 \\
    2023 & 1 minute  & No root & & \\
    2023 & 5 minutes & Unique & 0.0702 & 14.236 \\
    2023 & 10 minutes & Unique & 0.0961 & 10.402 \\
    2023 & 15 minutes & Unique & 0.0932 & 10.730 \\
    2024 & 1 minute  & Unique & 0.0413 & 24.218 \\
    2024 & 5 minutes & Unique & 0.1298 & 7.704 \\
    2024 & 10 minutes & Unique & 0.1018 & 9.826 \\
    2024 & 15 minutes & Unique & 0.1673 & 5.978 \\
    \hline
\end{longtable}
\endgroup

\clearpage

\subsection{Jump Diagnostics}
\label{app:jump_results}

Table~\ref{tab:jump_diagnostics_appendix} reports the share of returns classified as extreme at the intermediate threshold \(6\widehat{\sigma}_t\), together with their share of total squared variation.

\begin{table}[htbp]
    \centering
    \caption{Extreme-return diagnostics at the \(6\widehat{\sigma}_t\) threshold}
    \label{tab:jump_diagnostics_appendix}
    \small
    \begin{tabular}{ccc}
        \hline
        Year & Share of returns, \% & Share of squared variation, \% \\
        \hline
        2017 & 0.84 & 14.39 \\
        2018 & 0.35 & 10.09 \\
        2019 & 0.56 & 25.78 \\
        2020 & 0.35 & 28.64 \\
        2021 & 0.27 & 11.02 \\
        2022 & 0.57 & 21.38 \\
        2023 & 1.88 & 32.81 \\
        2024 & 0.58 & 18.65 \\
        \hline
    \end{tabular}
\end{table}

\clearpage

\subsection{Level Tests and Structural Breaks}
\label{app:level_break_results}

Table~\ref{tab:stationarity_break_appendix} reports the complete fixed-panel test statistics and break counts.

\begingroup
\scriptsize
\begin{longtable}{cclrrrr}
    \caption{Level tests and penalized break counts}
    \label{tab:stationarity_break_appendix}\\
    \hline
    Year & Frequency & ADF statistic & ADF \(p\) & KPSS statistic & KPSS \(p\) & Breaks \\
    \hline
    \endfirsthead
    \multicolumn{7}{c}{\tablename\ \thetable\ continued} \\
    \hline
    Year & Frequency & ADF statistic & ADF \(p\) & KPSS statistic & KPSS \(p\) & Breaks \\
    \hline
    \endhead
    2017 & 1 minute  & -36.759 & \(<0.001\) & 21.485 & 0.01 & 10 \\
    2017 & 5 minutes & -14.044 & \(<0.001\) & 10.279 & 0.01 & 10 \\
    2017 & 10 minutes & -10.149 & \(<0.001\) & 7.346 & 0.01 & 10 \\
    2017 & 15 minutes & -8.284 & \(<0.001\) & 6.218 & 0.01 & 10 \\
    2018 & 1 minute  & -43.002 & \(<0.001\) & 21.848 & 0.01 & 13 \\
    2018 & 5 minutes & -16.889 & \(<0.001\) & 11.220 & 0.01 & 8 \\
    2018 & 10 minutes & -11.353 & \(<0.001\) & 8.088 & 0.01 & 7 \\
    2018 & 15 minutes & -8.826 & \(<0.001\) & 6.858 & 0.01 & 6 \\
    2019 & 1 minute  & -43.778 & \(<0.001\) & 2.629 & 0.01 & 13 \\
    2019 & 5 minutes & -19.046 & \(<0.001\) & 1.433 & 0.01 & 7 \\
    2019 & 10 minutes & -13.104 & \(<0.001\) & 1.084 & 0.01 & 7 \\
    2019 & 15 minutes & -10.489 & \(<0.001\) & 0.933 & 0.01 & 7 \\
    2020 & 1 minute  & -30.988 & \(<0.001\) & 12.414 & 0.01 & 4 \\
    2020 & 5 minutes & -13.196 & \(<0.001\) & 6.292 & 0.01 & 4 \\
    2020 & 10 minutes & -9.989 & \(<0.001\) & 4.625 & 0.01 & 3 \\
    2020 & 15 minutes & -9.062 & \(<0.001\) & 3.909 & 0.01 & 3 \\
    2021 & 1 minute  & -36.185 & \(<0.001\) & 10.919 & 0.01 & 11 \\
    2021 & 5 minutes & -16.117 & \(<0.001\) & 5.704 & 0.01 & 10 \\
    2021 & 10 minutes & -10.593 & \(<0.001\) & 4.290 & 0.01 & 10 \\
    2021 & 15 minutes & -9.998 & \(<0.001\) & 3.743 & 0.01 & 10 \\
    2022 & 1 minute  & -41.512 & \(<0.001\) & 2.269 & 0.01 & 12 \\
    2022 & 5 minutes & -18.052 & \(<0.001\) & 1.244 & 0.01 & 11 \\
    2022 & 10 minutes & -13.737 & \(<0.001\) & 0.954 & 0.01 & 9 \\
    2022 & 15 minutes & -11.086 & \(<0.001\) & 0.826 & 0.01 & 9 \\
    2023 & 1 minute  & -38.602 & \(<0.001\) & 3.605 & 0.01 & 11 \\
    2023 & 5 minutes & -17.877 & \(<0.001\) & 2.072 & 0.01 & 11 \\
    2023 & 10 minutes & -14.108 & \(<0.001\) & 1.647 & 0.01 & 10 \\
    2023 & 15 minutes & -12.204 & \(<0.001\) & 1.489 & 0.01 & 9 \\
    2024 & 1 minute  & -34.567 & \(<0.001\) & 8.214 & 0.01 & 15 \\
    2024 & 5 minutes & -16.676 & \(<0.001\) & 4.467 & 0.01 & 13 \\
    2024 & 10 minutes & -12.332 & \(<0.001\) & 3.429 & 0.01 & 12 \\
    2024 & 15 minutes & -10.848 & \(<0.001\) & 3.023 & 0.01 & 7 \\
    \hline
\end{longtable}
\endgroup

\clearpage

\subsection{Scaling Diagnostics and Surrogate Comparisons}
\label{app:scaling_results}

Table~\ref{tab:scaling_appendix} reports the principal scaling statistics for all fixed year-frequency configurations. The final column records whether the upper endpoint of the 95 percent interval for wavelet \(c_2\) is below zero.

\begingroup
\scriptsize
\setlength{\tabcolsep}{3pt}
\begin{longtable}{cclrrrrl}
    \caption{Complete fixed-panel scaling diagnostics}
    \label{tab:scaling_appendix}\\
    \hline
    Year & Frequency & \(\Delta h\) & Shuffle \(p\) & Curvature \(c\) & Shuffle \(p\) & Wavelet \(c_2\) & Negative interval \\
    \hline
    \endfirsthead
    \multicolumn{8}{c}{\tablename\ \thetable\ continued} \\
    \hline
    Year & Frequency & \(\Delta h\) & Shuffle \(p\) & Curvature \(c\) & Shuffle \(p\) & Wavelet \(c_2\) & Negative interval \\
    \hline
    \endhead
    2017 & 1 minute  & 0.016 & 0.60 & 0.00091 & 0.05 & 0.027 & No \\
    2017 & 5 minutes & 0.017 & 1.00 & -0.00049 & 0.30 & 0.032 & No \\
    2017 & 10 minutes & 0.023 & 1.00 & -0.00344 & 0.05 & 0.033 & No \\
    2017 & 15 minutes & 0.046 & 1.00 & -0.00583 & 0.05 & 0.062 & No \\
    2018 & 1 minute  & 0.022 & 0.05 & -0.00055 & 0.10 & 0.029 & No \\
    2018 & 5 minutes & 0.019 & 1.00 & -0.00373 & 0.05 & -0.010 & No \\
    2018 & 10 minutes & 0.029 & 1.00 & -0.00506 & 0.05 & -0.008 & No \\
    2018 & 15 minutes & 0.053 & 0.65 & -0.00720 & 0.05 & 0.023 & No \\
    2019 & 1 minute  & 0.169 & 0.05 & -0.00326 & 0.05 & -0.064 & Yes \\
    2019 & 5 minutes & 0.209 & 0.05 & -0.00960 & 0.10 & -0.087 & Yes \\
    2019 & 10 minutes & 0.242 & 0.05 & -0.01058 & 0.05 & -0.094 & Yes \\
    2019 & 15 minutes & 0.239 & 0.05 & -0.01284 & 0.10 & -0.054 & Yes \\
    2020 & 1 minute  & 0.056 & 1.00 & 0.00196 & 0.05 & 0.003 & No \\
    2020 & 5 minutes & 0.041 & 1.00 & 0.00099 & 0.15 & -0.002 & No \\
    2020 & 10 minutes & 0.047 & 1.00 & -0.00161 & 0.15 & -0.001 & No \\
    2020 & 15 minutes & 0.072 & 1.00 & 0.00262 & 0.05 & 0.061 & No \\
    2021 & 1 minute  & 0.052 & 0.05 & -0.00067 & 0.10 & 0.040 & No \\
    2021 & 5 minutes & 0.060 & 0.95 & -0.00749 & 0.05 & -0.021 & No \\
    2021 & 10 minutes & 0.081 & 1.00 & -0.01035 & 0.05 & -0.020 & No \\
    2021 & 15 minutes & 0.086 & 1.00 & -0.01563 & 0.05 & 0.057 & No \\
    2022 & 1 minute  & 0.102 & 0.05 & -0.00285 & 0.05 & -0.040 & Yes \\
    2022 & 5 minutes & 0.117 & 0.05 & -0.01108 & 0.05 & -0.082 & Yes \\
    2022 & 10 minutes & 0.149 & 0.05 & -0.01227 & 0.05 & -0.081 & Yes \\
    2022 & 15 minutes & 0.156 & 0.05 & -0.01293 & 0.05 & -0.020 & No \\
    2023 & 1 minute  & 0.166 & 0.05 & -0.00624 & 0.05 & -0.136 & Yes \\
    2023 & 5 minutes & 0.176 & 0.05 & -0.01426 & 0.05 & -0.149 & Yes \\
    2023 & 10 minutes & 0.199 & 0.05 & -0.01591 & 0.05 & -0.166 & Yes \\
    2023 & 15 minutes & 0.214 & 0.05 & -0.01447 & 0.05 & -0.094 & Yes \\
    2024 & 1 minute  & 0.158 & 0.05 & -0.00361 & 0.05 & -0.037 & Yes \\
    2024 & 5 minutes & 0.138 & 0.05 & -0.01389 & 0.05 & -0.063 & Yes \\
    2024 & 10 minutes & 0.159 & 0.05 & -0.01742 & 0.05 & -0.066 & Yes \\
    2024 & 15 minutes & 0.183 & 0.05 & -0.01619 & 0.05 & -0.004 & No \\
    \hline
\end{longtable}
\endgroup

Table~\ref{tab:surrogate_probabilities_appendix} gives the complete empirical probabilities for the detailed five-minute surrogate analysis. Because 19 replications are used, a value of 0.05 indicates that none of the surrogate statistics equals or exceeds the observed statistic.

\begin{table}[htbp]
    \centering
    \caption{Five-minute empirical probabilities by surrogate type}
    \label{tab:surrogate_probabilities_appendix}
    \scriptsize
    \setlength{\tabcolsep}{3.5pt}
    \begin{tabular}{ccccccc}
        \hline
        Year &
        \multicolumn{2}{c}{Shuffled} &
        \multicolumn{2}{c}{Phase-randomized} &
        \multicolumn{2}{c}{IAAFT} \\
        \cline{2-7}
        & MF-DFA & Curvature & MF-DFA & Curvature & MF-DFA & Curvature \\
        \hline
        2017 & 1.00 & 0.30 & 0.30 & 0.15 & 0.90 & 1.00 \\
        2018 & 1.00 & 0.05 & 0.05 & 0.05 & 0.90 & 0.85 \\
        2019 & 0.05 & 0.10 & 0.05 & 0.05 & 0.05 & 0.30 \\
        2020 & 1.00 & 0.20 & 0.05 & 0.05 & 0.85 & 1.00 \\
        2021 & 1.00 & 0.05 & 0.05 & 0.05 & 0.45 & 1.00 \\
        2022 & 0.05 & 0.05 & 0.05 & 0.05 & 0.05 & 0.55 \\
        2023 & 0.05 & 0.05 & 0.05 & 0.05 & 0.05 & 0.05 \\
        2024 & 0.05 & 0.05 & 0.05 & 0.05 & 0.10 & 0.95 \\
        \hline
    \end{tabular}
\end{table}

\clearpage

\subsection{Temporal Trend Estimates}
\label{app:trend_results}

Table~\ref{tab:roughness_trends_appendix} reports the temporal tests for annual rolling root rates and median finite \(\widehat{H}\). The \(N=8\) observations at each frequency correspond to the eight start years.

\begin{table}[htbp]
    \centering
    \caption{Temporal trends in rolling roughness summaries}
    \label{tab:roughness_trends_appendix}
    \scriptsize
    \setlength{\tabcolsep}{3.5pt}
    \begin{tabular}{llcrr}
        \hline
        Frequency & Metric & Theil-Sen slope [95\% CI] & Spearman \(\rho\) & \(p\) \\
        \hline
        1 minute & Root rate & \(0.0000\;[-0.0625,\,0.0769]\) & 0.012 & 0.977 \\
        1 minute & Median \(\widehat{H}\) & \(0.0027\;[-0.0025,\,0.0062]\) & 0.310 & 0.456 \\
        5 minutes & Root rate & \(0.0193\;[-0.0278,\,0.0833]\) & 0.506 & 0.201 \\
        5 minutes & Median \(\widehat{H}\) & \(0.0054\;[-0.0019,\,0.0131]\) & 0.619 & 0.102 \\
        10 minutes & Root rate & \(0.0000\;[0.0000,\,0.0000]\) & 0.109 & 0.797 \\
        10 minutes & Median \(\widehat{H}\) & \(0.0058\;[-0.0008,\,0.0149]\) & 0.714 & 0.047 \\
        15 minutes & Root rate & \(0.0000\;[-0.0833,\,0.0208]\) & -0.026 & 0.951 \\
        15 minutes & Median \(\widehat{H}\) & \(0.0059\;[0.0006,\,0.0112]\) & 0.833 & 0.010 \\
        \hline
    \end{tabular}
\end{table}

Table~\ref{tab:scaling_trends_appendix} reports the corresponding temporal tests for the three scaling diagnostics in the fixed panel. Each frequency-specific estimate uses eight yearly observations.

\begin{table}[htbp]
    \centering
    \caption{Temporal trends in fixed-window scaling diagnostics}
    \label{tab:scaling_trends_appendix}
    \scriptsize
    \setlength{\tabcolsep}{3.5pt}
    \begin{tabular}{llcrr}
        \hline
        Frequency & Diagnostic & Theil-Sen slope [95\% CI] & Spearman \(\rho\) & \(p\) \\
        \hline
        1 minute & MF-DFA width & \(0.020150\;[-0.002209,\,0.036435]\) & 0.595 & 0.120 \\
        1 minute & Log-moment curvature & \(-0.000749\;[-0.002407,\,0.000135]\) & -0.714 & 0.047 \\
        1 minute & Wavelet \(c_2\) & \(-0.012057\;[-0.045535,\,0.005355]\) & -0.452 & 0.260 \\
        5 minutes & MF-DFA width & \(0.019133\;[-0.008447,\,0.044743]\) & 0.619 & 0.102 \\
        5 minutes & Log-moment curvature & \(-0.002010\;[-0.003720,\,-0.000494]\) & -0.786 & 0.021 \\
        5 minutes & Wavelet \(c_2\) & \(-0.015461\;[-0.049134,\,0.004124]\) & -0.619 & 0.102 \\
        10 minutes & MF-DFA width & \(0.023398\;[-0.010642,\,0.050675]\) & 0.619 & 0.102 \\
        10 minutes & Log-moment curvature & \(-0.001958\;[-0.003638,\,-0.001331]\) & -0.810 & 0.015 \\
        10 minutes & Wavelet \(c_2\) & \(-0.017065\;[-0.054850,\,0.004321]\) & -0.619 & 0.102 \\
        15 minutes & MF-DFA width & \(0.020673\;[-0.006310,\,0.047334]\) & 0.619 & 0.102 \\
        15 minutes & Log-moment curvature & \(-0.001447\;[-0.003503,\,-0.000031]\) & -0.786 & 0.021 \\
        15 minutes & Wavelet \(c_2\) & \(-0.010425\;[-0.051637,\,0.011112]\) & -0.548 & 0.160 \\
        \hline
    \end{tabular}
\end{table}

\clearpage
\bibliographystyle{plainnat}
\bibliography{references}

\end{document}